\documentclass[conference]{IEEEtran}
\IEEEoverridecommandlockouts
\usepackage{cite}
\usepackage{amsmath,amssymb,amsfonts}
\usepackage{algorithmic}
\usepackage{graphicx}
\usepackage{textcomp}
\usepackage{epsfig}
\usepackage{epstopdf}
\usepackage{xcolor}
\def\BibTeX{{\rm B\kern-.05em{\sc i\kern-.025em b}\kern-.08em
    T\kern-.1667em\lower.7ex\hbox{E}\kern-.125emX}}

\def\be{\begin{equation}}
\def\ee{\end{equation}}

\def\mih{\mathbf{h}}

\def\mix{\mathbf{x}}

\def\mi0{\mathbf{0}}

\begin{document}

\title{ Cooperative Artificial Neural Networks for Rate-Maximization in Optical Wireless Networks
{\footnotesize \textsuperscript{}}
 
 \thanks{This work  has been supported in part by the Engineering and Physical Sciences Research Council (EPSRC), in part by the INTERNET project under Grant EP/H040536/1, and in part by the STAR project under Grant EP/K016873/1 and in part by the TOWS project under Grant EP/S016570/1. All data are provided in full in the results section of this paper.}
}

\author{\IEEEauthorblockN{Ahmad Adnan Qidan, Taisir El-Gorashi, Jaafar M. H. Elmirghani}
\IEEEauthorblockA{School of Electronic and Electrical Engineering, University of Leeds, LS2 9JT, United Kingdom 
\\Email: \{a.a.qidan, t.e.h.elgorashi,j.m.h.elmirghani\}@leeds.ac.uk}


}

\maketitle

\maketitle

\begin{abstract}
Recently, Optical wireless communication (OWC) have been considered  as a key element in the next generation of wireless communications due to its potential in supporting unprecedented  communication speeds. In this paper,  infrared lasers referred to as vertical-cavity surface-emitting lasers (VCSELs) are used as transmitters sending  information to multiple users. In OWC, rate-maximization optimization problems are usually complex due to the high number of optical access points (APs) needed to ensure coverage. Therefore, practical solutions with low computational time  are essential to cope with frequent updates in user-requirements that might occur. In this context, we formulate an optimization problem to determine the optimal user association and resource allocation in the network, while the serving time is partitioned into a series of time periods. Therefore, cooperative ANN models are designed  to estimate and predict the association and resource allocation variables for each user such that sub-optimal solutions can be obtained within a certain period of time prior to its actual starting, which makes the solutions valid and in accordance with the demands of the users at a given time. The results show the effectiveness of the proposed model in maximizing the sum rate of the network compared with counterpart models. Moreover, ANN-based solutions are close to the optimal ones with low computational time.

\end{abstract}

\IEEEpeerreviewmaketitle

\section{Introduction} 
The evolution of Internet-based technologies in recent days has led to challenges in terms of traffic congestion and lack of resources and secrecy that current wireless networks have failed to support. 
Therefore, optical wireless communication (OWC) has attracted  massive interest from scientific researchers to provide unprecedented communication speeds. Basically, OWC sends information modulated on the optical band, which offers huge license free-bandwidth and high spectral and energy efficiency.
In \cite{6011734}, light-emitting diodes (LEDs) were used as transmitters providing  data rates in gigabit-per-second (Gbps) communication speeds. Despite the characteristics of  LEDs, the modulation speed is limited, and they are usually deployed for providing illumination, and therefore, increasing the number of transmitters must be in compliance with the recommended illumination levels in such indoor environments.
Alternatively, infrared  lasers such as vertical-cavity surface-emitting lasers (VCSELs) were used in \cite{9803253} to serve users at Terabit-per-second (Tbps) aggregate data rates, which makes OWC as a strong candidate in the next generation of wireless communications. However, the transmit power of the VCSEL can be harmful to human eyes if it  operates at high power levels without considering eye safety regulations. 

Optimization problems for rate-maximization were formulated in \cite{9685357,JLT1111} to enhance  the  spectral efficiency of OWC networks. In particular, a resource allocation approach was designed in \cite{9685357} to guarantee high quality of service for users with different demands. In \cite{JLT1111}, centralized and decentralized algorithms were proposed to  maximize the sum rate of the network under the capacity constraint of the optical AP. It is worth pointing out that optimization problems in the context of rate-maximization  are usually defined as complex problems that are time consumers. Recently, machine learning (ML) techniques have been considered to provide practical solutions for NP-hard optimization problems. In \cite{df}, a deep learning algorithm was used for power allocation in massive multiple-input multiple-output (MIMO) to achieve relatively high spectral efficiency at low loss. In \cite{9839259},  an artificial neural network (ANN) model  was trained for resource allocation-based rate maximization in OWC network. It is shown that a closed form solution to the optimum solution of exhaustive search  can be achieved at low complexity. However, the use of ML techniques in optical or RF wireless networks is still under investigation especially  in complex scenarios where decisions for example in rate-maximization  must be made promptly.  

In contrast to the work in the literature, in this paper, we design two ANN models working in cooperation  to maximize the sum rate of a discrete-time OWC network in which the serving time is partitioned into consecutive periods of time. First, a multi user OWC system model is defined  where a transmission scheme referred to as blind interference alignment (BIA) is applied for multiple access services. Then, an optimization problem is formulated to find the optimum user-association and resource allocation during a certain period of time. The computational time of solving such complex optimization problems exceeds the time during which the optimum solution must be determined. Therefore, two ANN models are designed and trained to maximize the sum rate of the network during the intended period of time prior to its staring by exploiting the records of the network in the previous period of time and performing prediction. The results show the ability of the trained ANN models in providing accurate solutions close to the optimum ones.
\begin{figure}[t]
\begin{center}\hspace*{0cm}
\includegraphics[width=0.83\linewidth]{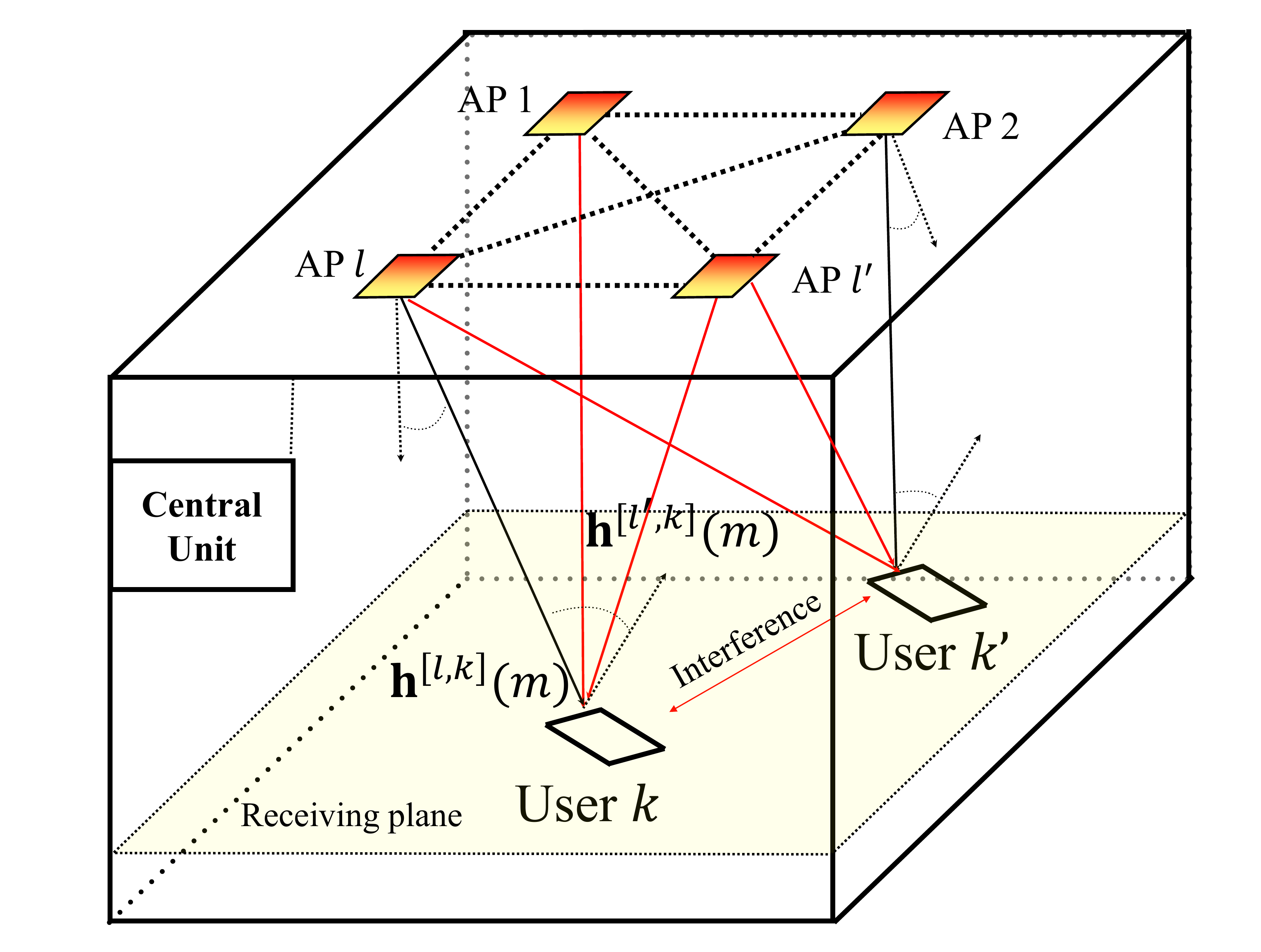}
\end{center}
\vspace{-2mm}
\caption{ An OWC system with $ L $ optical APs serving $ K $ users. }\label{Figmodel}
\vspace{-2mm}
\end{figure}

\section{System Model}
\label{sec:system}
We consider a discrete-time downlink  OWC network as shown in Fig. \ref{Figmodel}, where multiple optical APs given by $ L $, $ l=\{1, \dots, L\} $, are deployed on the ceiling to serve multiple users given  by $ K $, $ k=\{1, \dots, K\} $, distributed on the communication floor. Note that, the VCSEL  is  used as a transmitter, and therefore, each optical AP consists  of   $ L_{v} $ VCSELs to extend  its coverage area. On the user side, a  reconfigurable optical detector  with $ M $ photodiodes providing a wide field of view (FoV)\cite{8636954} is used to ensure that each user has more than one optical link available at a given time. In this work, the serving time in the network is partitioned into a set of time periods given by $\mathcal{T}$, where $ t=\{ 1, \dots, t, t+1, \dots \mathcal{T}\} $, and the duration of each time period is $ \tau $. In this context,  the signal received by a generic user $ k $, $ k \in K $,  connected to AP $ l $ during the period of time $t+1$ can be expressed as 
\begin{equation}
y^{[l,k]}(t+1)=\mih_{t+1}^{[l,k]}(m^{[l,k]})^{T} \mix(t+1)+  z^{[l,k]}(t+1),
\end{equation}
where $ m \in M $ is a photodiode of user $ k $,   $ \mih_{t+1}^{[l,k]}(m^{[l,k]})^{T} \in \mathbb{R}_+^{l\times 1} $, is the channel matrix, $  \mix(t+1) $ is the transmitted signal, and $ z^{[l,k]}(t+1) $ is real valued additive white Gaussian noise with zero mean and variance given by the sum of shot noise,
thermal noise and the intensity noise of the laser.
In this work, all the optical APs are connected through a central unit (CU) to exchange essential information for solving optimization problems. It is worth mentioning that the distribution of the users  is known at the central unite, while the channel state information (CSI) at the transmitters is limited to the channel coherence time due to the fact that BIA is implemented  for  interference management \cite{8636954,Gou11}. 
\subsection{Transmitter}
The VCSEL transmitter has Gaussian beam profile with multiple modes. For lasers, the power distribution is determined based on the beam waist $ W_{0} $, the wavelength $ \lambda $ and the distance $ d $ between the transmitter and  user. Basically, the beam radius of the VCSEL at photodiode $ m $ of user $ k $ located on the communication floor at distance $d$ is given by 
\begin{equation}
W_{d}=W_{0} \left( 1+ \left(\frac{d}{d_{Ra}}\right)^{2}\right)^{1/2},
\end{equation}
where $ d_{Ra} $ is the Rayleigh range. Moreover, the spatial distribution of the intensity of VCSEL transmitter $ l $ over the transverse plane at distance $ d $ is given by 
\begin{equation}
I_{l}(r,d) = \frac{2 P_{t,l}}{\pi W^{2}_{d}}~ \mathrm{exp}\left(-\frac{2 r^{2}}{W^{2}_{d}}\right).
\end{equation}
Finally, the power received by photodiode $ m $  of user $ k $  from  transmitter  $ l $ is given by 
\begin{equation}
\label{power}
\begin{split}
&P_{m,l}=\\
&\int_{0}^{r_m} I(r,d) 2\pi r dr = P_{t,l}\left[1- \mathrm{exp} \left(\frac{ -2r^{2}_{m}}{W^{2}_{d}}\right)\right],
\end{split}
\end{equation}
where $ r_m $ is the radius of photodiode $ m $. Note that, $ A_m = \frac{A_{rec}}{M} $, $ m \in M $, is the detection area of photodiode  $ m $, assuming $ A_{rec} $ is the whole detection area of the receiver. In \eqref{power}, the location of  user $ k $ is considered right under transmitter $ l $, more details on the power calculations of the laser are in \cite{9803253}.
\subsection{Blind interference alignment}
\begin{figure}[t]
\begin{center}\hspace*{0cm}
\includegraphics[width=0.65\linewidth]{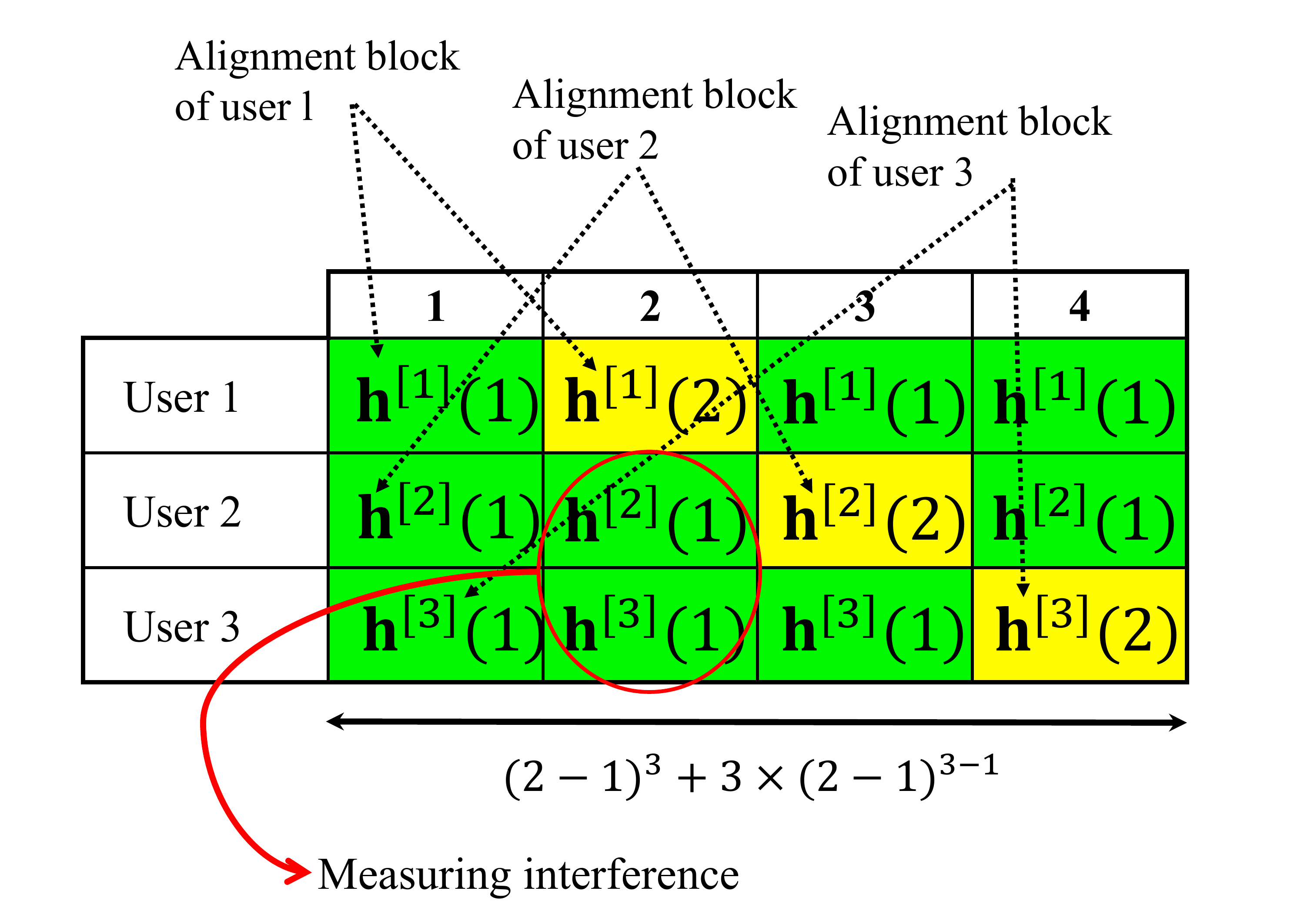}
\end{center}
\vspace{-2mm}
\caption{ Transmission block of BIA for a use case.}\label{bia}
\vspace{-2mm}
\end{figure}

BIA is a transmission scheme proposed for RF and optical networks to manage multi-user interference  with no CSI at the transmitters \cite{8636954,Gou11}, showing superiority over other transmit precoding schemes with CSI such as zero-forcing (ZF). Basically, the transmission block of BIA allocates  multiple alignments block to each user following a unique methodology. For instance, an AP with $ L_{v}=2  $ transmitters serving $ K=3 $ users, one alignment block is allocated to each user as shown in Fig. \ref{bia}. For the general case where an optical AP composed of  $ L_v $ transmitters serving $ K $ users, BIA allocates $ \ell= \big\{1, \dots, (L_{v}-1)^{K-1}\big\} $ alignment blocks to each user over a transmission block consisting of  $ (L_{v}-1)^{K}+K (L_{v}-1)^{K-1}  $ time slots. In this context,   user $  k$ receives the symbol $ \mathbf{u}_{\ell}^{[l,k]} $ from AP $ l $ during  the $ \ell $-th  alignment block as follows 
\begin{equation}
\label{recibia}
\mathbf{y}^{[l,k]} = \mathbf{H}^{[l,k]}\mathbf{u}_{\ell}^{[l,k]} +\sum_{ \substack{l' = 1,  l'\neq l}}^{L}
   \sqrt{\alpha_{l'}^{[l,k]}}\mathbf{H}^{[l',k]}\mathbf{u}_{\ell}^{[l',k]}+  \mathbf{z}^{[l,k]},
\end{equation}
where $ \mathbf{H}^{[l,c]} $ is the channel matrix of user $k$. It is worth mentioning that user $ k $ is equipped with a reconfigurable detector that has the ability to provide $L_v$ linearly independent channel responses, i.e.,  
\begin{equation}
\mathbf{H}^{[l,k]} = \begin{bmatrix} \mathbf{h}^{[l,k]}(1)& \mathbf{h}^{[l,k]}(2)& \dots & \mathbf{h}^{[l,k]}(L_v)
 \end{bmatrix} \in \mathbb{R}_+^{L_{v} \times 1}.
\end{equation}
In \eqref{recibia}, $ \alpha_{l'}^{[l,k]} $ is the signal-to-interference ratio (SIR) received at user $ k $ due to other APs $ l \neq l' $, and $ \mathbf{u}_{\ell}^{[l',k]} $
represents the interfering symbols received from the adjacent APs during the alignment block  $ \ell $ over which the desired symbol $ \mathbf{u}_{\ell}^{[k,c]} $ is received. It is worth pointing out that frequency reuse is usually applied to avoid inter-cell interference so that the interfering symbol  $ \mathbf{u}_{\ell}^{[l',k]} $ can be teared as noise.   
Finally, $\mathbf{z}^{[k,c]}$ is defined as noise resulting from interference subtraction, and it is given by a covariance matrix, i.e.,
\begin {equation}
\mathbf{R_{z_p}} = 
\begin{bmatrix} 
(K)\mathbf{I}_{L_{v}-1} & \mathbf{0}\\ 
\mathbf{0} & 1\\
\end{bmatrix}.
\end{equation}
According to \cite{8636954}, the BIA-based data rate received by user $k$ from its corresponding APs during the period of time $ (t+1) $ is expressed as
\begin{multline}
\label{rate}
r^{[l,k]} (t+1) =\\ B_{t+1}^{[l,k]} \mathbb{E}\left[\log\det\left(\mathbf{I}_{L_{v}} + P_{\rm{str}} \mathbf{H}_{t+1}^{[l,k]}{\mathbf{H}^{[l,k]}}^{H} \mathbf{R_{\tilde{z}}}^{-1}(t+1) \right)\right],
\end{multline}
where $B_{t+1}^{[l,k]}=\dfrac{(L_{v}-1)^{K-1}}{(L_{v}-1)^{K}+K(L_{v}-1)^{K-1}}= \dfrac{1}{K+L_{v}-1}$
is the ratio of the alignment blocks allocated to each user connected to AP $ l $ over the entire transmission block, $ P_{\rm{str}} $ is the power allocated to each stream and
\begin{equation}
 \mathbf{R_{\tilde{z}}}(t+1)= \mathbf{R_{z_p}} + P_{\rm{str}}\sum_{l' = 1}^{L} \alpha_{l'}^{[l,k]}\mathbf{H}_{t+1}^{[l',k]}{\mathbf{H}^{[l',k]}}^{H},
\end{equation}
is the covariance matrix of the noise plus  interference received from other APs $ l \neq l'$.
\section{Problem Formulation}
We formulate an optimization problem in a discrete time OWC
system aiming to maximize  the sum rate of the users by determining  the optimum user assignment and resource allocation simultaneously. 
It is worth mentioning that data rate maximization in the network must be achieved during each period of time $ t$, otherwise, it cannot be considered as a valid solution  due to the fact that user-conditions are subject to changes in the next period of time. Focussing on period of time $ t+1 $,  the utility function of sum rate maximization is  given by        
\begin{equation}
U(x,e)=  \sum_{k \in K} \varphi \left( \sum_{ l\in L} x^{[l,k]}_{t+1} R^{[l,k]}({t+1})\right),
\end{equation}
where $ x^{[l,k]}_{t+1} $ is an assignment variable that determines the connectivity of  user $ k $ to optical AP $  l$, where $ x^{[l,k]}_{t+1}=1 $ if user $ k $ is assigned to AP $ l $ during the period of time $ t+1 $, otherwise, it equals 0. Moreover,  the actual data rate of user $ k $ during ${t+1} $ is $ R^{[l,k]}({t+1})=  e^{[l,k]}_{t+1} r^{[l,k]}({t+1}) $, where $  e^{[l,k]}_{t+1} $, $ 0\leq e^{[l,k]}_{t+1} \leq 1$, determines the resources devoted  from AP $ l $ to serve user $ k $, and $ r^{[l,k]}({t+1}) $ is the user rate given by equation \eqref{rate}.  The sum rate maximization during the period of time $ {t+1} $ can be obtained by solving the optimization problem as follows
\begin{equation}
\label{pro-c1}
\begin{aligned} 
\mathbf{P1:} ~~\max_{x,e} \quad \sum_{k \in K} \varphi \left( \sum_{ l\in L} x^{[l,k]}_{t+1} R^{[l,k]}({t+1})\right)\\
\textrm{s.t.} \quad  \sum\limits_{l\in L}   x^{[l,k]}_{t+1}=1,~~~~~~~~~~~\forall k\in K \\
\quad  \sum\limits_{k\in K}  x^{[l,k]}_{t+1} R^{[l,k]}({t+1}) \leq \rho_{l},~~~~~\\~~~~~~~~\forall k\in K\\
  \quad  R_{min} \leq x^{[l,k]}_{t+1} R^{[l,k]}({t+1}) \leq R_{max}, \\~~~~~~~~\forall l\in L, k\in K \\
x^{[l,k]}_{t+1} \in \big\{0,1\big\}, l \in L, k \in K,~~~~~~~\\
\end{aligned}
\end{equation}
where $ \varphi (.) = \log(.) $ is a logarithmic function that achieves proportional fairness among users \cite{879343}, and $ \rho_{l} $ is the capacity limitation of AP $  l$ . The first constraint in \eqref{pro-c1} guarantees that each user is assigned to only one AP,  while the second constraint ensures that each AP is not overloaded. Moreover, the achievable user rate must be within a certain range as in the third constraint where   $  R_{min} $ is the minimum  data rate required by a given user and  $  R_{max} $ is the maximum  data rate that user $ k $ can receive. It is worth mentioning that imposing  the  third constraint helps in minimizing the waste of the resources  and guarantees high quality of service. Finally, the last constraint defines the feasible region of the optimization problem. 

The optimization problem in \eqref{pro-c1} is defined is as a  mixed integer non-linear
programming (MINLP) problem in which  two variables, $ x^{[l,k]}_{t+1} $ and $ e^{[l,k]}_{t+1} $, are coupled. 
Interestingly, some of the deterministic algorithms can be used to solve such  complex MINLP problems with high computational time. However, the application of these algorithms  in real scenarios is not practical to solve optimization problems like  in \eqref{pro-c1}, where the optimal solutions must be determined within a certain period of time.
One of the assumptions for relaxing the main optimization problem in \eqref{pro-c1} is to connect each user to more than one AP, which means that the association variable $ x^{[l,k]}_{t+1} $ equals to 1. In this context, the optimization problem can be rewritten as   
\begin{equation}
\label{pro-c2}
\begin{aligned} 
\mathbf{P2:} ~~\max_{e} \quad \sum_{k \in K} \varphi \left( \sum_{ l\in L} e^{[l,k]}_{t+1} r^{[l,k]}({t+1})\right)\\
\textrm{s.t.} \quad  \sum\limits_{k\in K} e^{[l,k]}_{t+1} r^{[l,k]}({t+1}) \leq \rho_{l},~~~~~\\~~~~~~~~\forall k\in K\\
  \quad  R_{min} \leq e^{[l,k]}_{t+1} r^{[l,k]}({t+1}) \leq R_{max}, \\~~~~~~~~\forall l\in L, k\in K \\
0\leq e^{[l,k]}_{t+1} \leq 1, l \in L, k \in K.~~~~~~~\\
\end{aligned}
\end{equation}
Note that,  considering our assumption of full connectivity,  the variable  $  x^{[l,k]}_{t+1} $ is eliminated. Interestingly, the optimization  problem in \eqref{pro-c2} can be solved in a distributed manner on the AP and user sides using  the full decomposition method via Lagrangian multipliers \cite{FRLHANZO}. Thus, the Lagrangian function is

\begin{multline}
\label{eq:lag}
 f\left(e,\mu, \xi_{\max},\lambda_{\min} \right) =\sum_{ k\in K} \sum_{ l\in L} \varphi \left( e^{[l,k]}_{t+1} r^{[l,k]}({t+1})\right)+\\  \sum_{l\in L} \mu^{[l]}_{t+1}
\left(\rho_{l}- \sum_{k\in K} R^{[l,k]}({t+1})\right)+
 \sum_{k\in K} \xi^{[k]}_{{t+1},\max}
 \\ \left(R_{max}- \sum_{l\in \L}  R^{[l,k]}({t+1})\right) +\sum_{k \in K} \lambda^{[k]}_{{t+1},\min} \\ \left( \sum_{l\in L} R^{[l,k]}({t+1})- R_{min}\right),
\end{multline}
where $ \mu$, $\xi_{\max} $ and $\lambda_{\min}$ are the Lagrange multipliers according  to the first and second constraints in \eqref{eq:lag}, respectively.
However, the assumption of users assigned to more than one AP is unrealistic in real time scenarios where users might not see more than one AP at a given time due to blockage. Therefore, focusing on resource allocation more than user association  as in \eqref{pro-c2} can cause relatively high waste of resources due to the fact that an AP might allocate resources to users blocked from receiving its LoS link. In the following, an alternative solution is proposed using ANN models.  
\subsection{Artificial neural network }
Our aim in \eqref{pro-c1} is to provide optimal solutions during the period of time $ t+1 $. Therefore, our ANN model  must have the ability  to exploit the solutions of the optimization problem in the previous  period of time $ t $.
Given that, the problem in hand can be defined as time series prediction, 
 Focussing on the optimization problem in \eqref{pro-c1}, calculating  the network assignment vector $ \mathbf {X}_{t+1} $ involves high complexity. Therefore, having an ANN model that is able  to perform prediction for the network assignment vector can considerably minimize the computational time, while valid sub-optimum solutions are obtained within a certain period of time. As in \cite{9839259}, we design a convolutional neural network (CNN)  to estimate the network assignment vector  denoted by $  \widehat{{\mathbf{X}}}_{t}  $ during a given period of time $ t $ based on user-requirements sent to the whole set of the APs through uplink transmission. It is worth mentioning that the CNN model must be trained over a data set generated from solving the original optimization problem as in the following  sub-section. For prediction, we consider the use of long-short-term-memory (LSTM) model classified as a recurrent neural network (RNN)\cite{7960065}, which is known to solve complex sequence problems through time. Once, the network assignment vector is estimated during the period of time $ t $, it is fed into the input layer of the LSTM model trained  to predict the network assignment vector $ \widetilde{{\mathbf{X}}}_{t+1} $ during the next period of time $ t+1 $ prior to its starting. Note that, resource allocation can be performed in accordance with the predicted network assignment vector to achieve data rate maximization during the intended period of time.
\subsection{Offline phase}
We first train the CNN  model over a dataset with $ N $ size within each period of time to determine  an accurate set of weight terms that can perfectly map between the information sent to the input layer of the ANN model and its output layer. Note that, the CNN model  aims to estimate the network assignment vector at a given time. For instance during  period of time $ t $, the CNN model provides an estimated network assignment vector within the interval $ [\widehat{{\mathbf{X}}}_{t-\tau+1}, \widehat{{\mathbf{X}}}_{t-\tau+2}, \dots, \widehat{{\mathbf{X}}}_{t}  ] $, which then can be fed into the input layer of the LSTM model to predict the network assignment vector  $ \widetilde{{\mathbf{X}}}_{t+1} $. In this context, the CNN  model must be trained during the period of time $ t $ over data points generated  from solving the following problem
\begin{equation}
\label{pro-3}
\begin{aligned} 
\mathbf{P3:} ~~\max_{x} \quad \sum_{k \in K} \varphi \left( \sum_{ l\in L} x^{[l,k]}_{t} \frac{r^{[l,k]}({t})}{\sum_{k \in K} x^{[l,k]}_{t}}\right)\\
\textrm{s.t.} \quad  \sum\limits_{l\in L}   x^{[l,k]}_{t}=1,~~~~~~~~~~~\forall k\in K \\
x^{[l,k]}_{t} \in \big\{0,1\big\}, l \in L, k \in K.~~~~~~~\\
\end{aligned}
\end{equation}
This  optimization problem  is a rewritten form of the problem in \eqref {pro-c1} with the assumption of uniform resource allocation, i.e.,  $ e^{[l,k]}_{t}=\frac{1}{K_{l}}  $, where $ K_{l}= \sum_{k \in K} x^{[l,k]}_{t} $. It is worth pointing out that this assumption is considered  due to the fact that once the estimation and prediction processes for the network assignment vector  are done using CNN and LSTM models, respectively, resource allocation is performed at each optical AP to satisfy the requirements of the users as in sub-section \ref{sub}.  The optimization problem  in \eqref {pro-3} can be solved through brute force search with a complexity that increases exponentially with the size of the network.
Note that, since the dataset is generated in an offline phase, complexity is not an issue.

For the LSTM model, the dataset is generated over $ \mathcal{T} $ consecutive period of times. Then, it is processed to train the LSTM model for  determining a set of wight terms that can accurately predict the network assignment vector during a certain period of time. Interestingly, the training of the LSTM model for predicting  $ \widetilde{{\mathbf{X}}}_{t+1} $ during $ t+1 $ is occurred over date points  included in the dataset during the previous time duration $ \tau $, i.e., $ [\widehat{{\mathbf{X}}}_{t-\tau+1}, \widehat{{\mathbf{X}}}_{t-\tau+2}, \dots, \widehat{{\mathbf{X}}}_{t}] $.
 

\subsection{Online application}
\label{sub}
After generating the dataset and training the ANN models in an offline phase, their application is considered  at the optical APs to perform instantaneous  data rate-maximization during a certain period of time $ t+1 $  by finding the optimum user association and resource allocation. Basically, the users send their requirements to the optical APs at the beginning of the period of time $ t $ through uplink transmission. Subsequently, these information are injected into the trained CNN model  to estimate the network assignment vector $ \widehat{{\mathbf{X}}}_{t} $ during the  interval $ [t-\tau+1, t-\tau+2, \dots, t] $, which then can be used as information for the  input layer of the LSTM trained to predict the network assignment  vector $ \widetilde{{\mathbf{X}}}_{t+1} $ during the next period of time  prior to its  actual starting. Once the  network assignment variable $ x^{[l,k]}_{t+1} $ is predicted for each user $ k $ during $ t+1 $, resource allocation is determined at each AP according to equation \eqref{eq:lag} as follows
\begin{multline}
\label{OPT4}
\mathcal{L}(e,\mu, \xi_{\max},\lambda_{\min})=\\ \sum_{k \in K_{l}} \varphi \left( e^{[l,k]}_{t+1} r^{[l,k]}({t+1})\right)- \mu^{[l]}_{t+1}  \sum_{k \in K_{l}}e^{[l,k]}_{t+1} r^{[l,k]}({t+1})
\\-\sum_{k \in K_{l}} ( \xi^{[k]}_{{t+1},\max}-\lambda^{[k]}_{{t+1},\min}) ~ e^{[l,k]}_{t+1} r^{[l,k]}({t+1}).
\end{multline}
 The optimum resources  allocated to  user $ k $ associated with AP $ l $  during ${t+1}$ is determined by taking the partial derivative of $ \mathcal{L}(e,\mu, \xi_{\max},\lambda_{\min})  $ with respect to $ e^{[l,k]}_{t+1} $. Therefore, it is given by

\begin{multline}
\label{OPT5}
\left(\dfrac{  \partial \varphi \left( e^{[l,k]}_{t+1} r^{[l,k]}({t+1})\right)} {\partial e_{t+1}}\right)= \\r^{[l,k]}({t+1}) \left(\mu^{[l]}_{t+1}+ \xi^{[k]}_{{t+1},\max}-\lambda^{[k]}_{{t+1},\min}\right), 
\end{multline}
Otherwise, considering the definition of $ \dfrac{\partial \mathcal{L} \left(e\right)} {\partial e} $ as monotonically decreasing function with $ e^{[l,k]}_{t+1} $, 
 the partial derivative $ \dfrac{\partial \mathcal{L} \left(e\right)} {\partial e} \vert _{e^{[l,k]}_{t+1}=0}\leq0 $ means that the optimum value $ e^{*[l,k]}_{t+1} $ equals zero, while  $ \dfrac{\partial \mathcal{L}\left(e\right)} {\partial e} \vert _{e^{[l,k]}_{t+1}=1} \geq 0 $ means that the optimum value $ e^{*[l,k]}_{t+1} $ equals one.  At this point, the gradient projection method is applied to solve the dual problem,  and the Lagrangian multipliers in \eqref{OPT4} are updated as follows 

\begin{equation}
\label{var}
\mu^{[l]}_{t+1}(i)= \left[\mu^{[l]}_{t+1}(i-1)-\Omega_{\varepsilon} \left(\alpha_l-\sum_{k\in K_{l}} R^{[l,k]}({t+1})\right) \right]^{+},
\end{equation}

\begin{equation}
\label{nu}
\xi^{[k]}_{{t+1},\max}(i)= \Bigg[\xi^{[k]}_{{t+1}}(i-1)-\Omega_{\nu}\Bigg(R_{\max}-R^{[l,k]}({t+1})\Bigg)\Bigg]^{+},
\end{equation}

\begin{equation}
\label{lam}
\lambda^{[k]}_{{t+1},\min} (i)=\Bigg[\lambda^{[k]}_{{t+1}}(i-1)-\Omega_{\lambda}\Bigg( R^{[l,k]}({t+1})-R_{\min}\Bigg)\Bigg]^{+},
\end{equation}
where $ i $ denotes the iteration of the gradient algorithm and [:]+ is a projection on the positive orthant. The Lagrangian variables  work as  indicators between the users and APs to maximize the sum rate of the network, while ensuring that each AP is not overloaded and the users receiver their demands. Note that, the resources are determined based on the  predicted network assignment vector $ \widetilde{{\mathbf{X}}}_{t+1} $. Therefore, at the beginning of the period of time  $ t+1 $,  each AP sets its link price according to  \eqref{var}, and the users update and broadcast their demands as in \eqref{nu} and \eqref{lam}. These values remain fixed  during the whole time interval so that the trained CNN estimate a new  assignment vector  to feed the LSTM model in order to predict $ \widetilde{{\mathbf{X}}}_{t+1} $ for the next period of time $ t+2  $. 

\begin{table}
\centering
\caption{Simulation Parameters}
\begin{tabular}{|l|c|}
\hline
{\bf Laser-based OWC parameter}	& {\bf Value} \\ \hline\hline
Laser Bandwidth	& 5 GHz \\\hline
Laser Wavelength  & 830 nm \\\hline
Laser beam waist & $ 5~ \mu $m \\\hline
Physical area of the photodiode	&15 $\text{mm}^2$ \\\hline
Receiver FOV & 45 deg \\\hline
Detector responsivity 	& 0.9 A/W \\\hline
Gain of optical filter & 	1.0 \\\hline
Laser noise	& $-155~ dB/H$z  \\\hline\hline
{\bf ANNs parameter}	& {\bf Value} \\ \hline\hline
 Model & CNN  and LSTM \\\hline
Dataset size & $ 5000-10000 $\\\hline
Training & $ 90\% $ of dataset\\\hline
Validation & $ 10\% $ of dataset\\\hline
\end{tabular}
\end{table}

\begin{figure}[t]
\begin{center}\hspace*{0cm}
\includegraphics[width=0.91\linewidth]{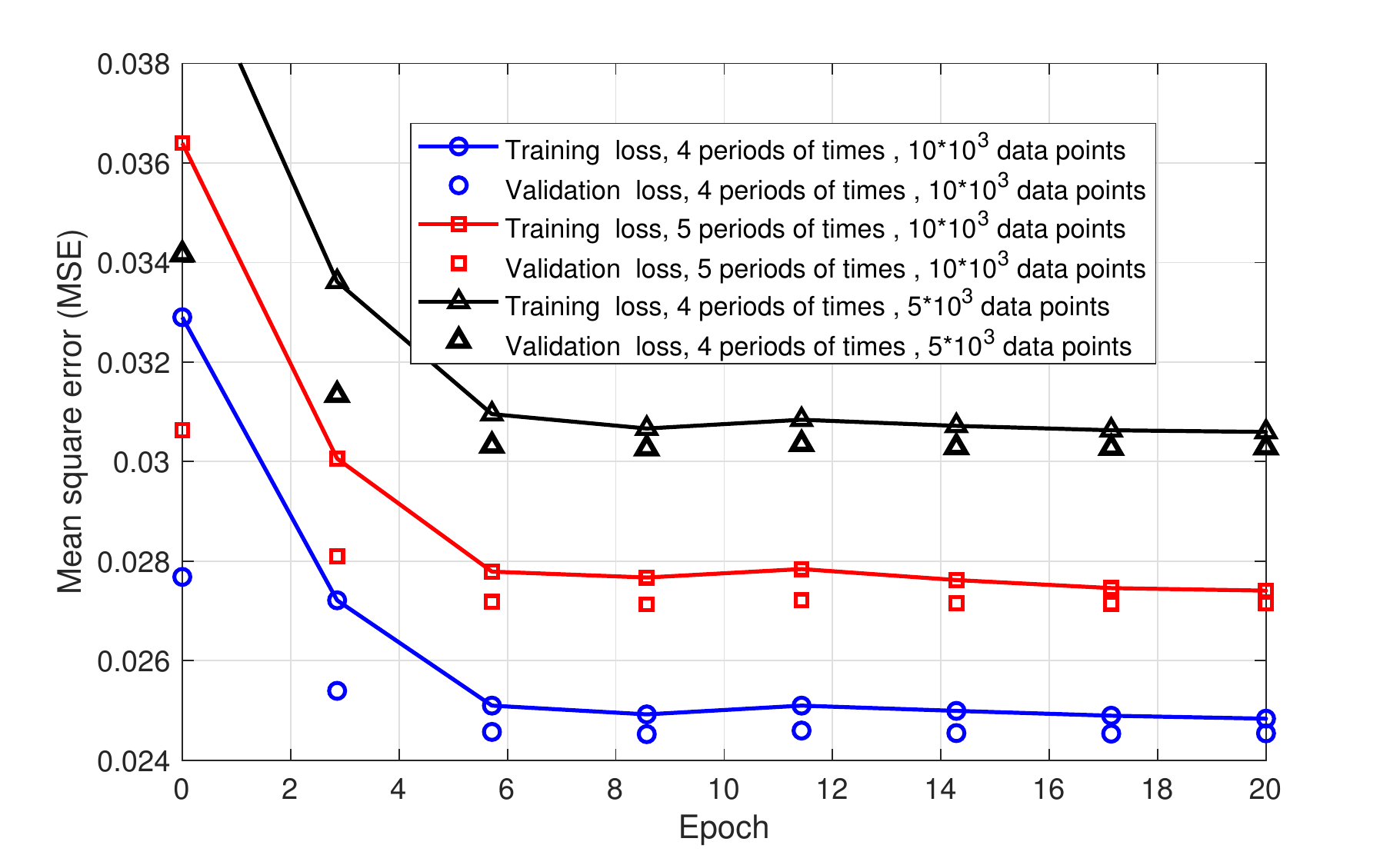}
\end{center}
\vspace{-2mm}
\caption{ The performance of the ANN model trained for prediction. }\label{re1}
\vspace{-2mm}
\end{figure}

\begin{figure}[t]
\begin{center}\hspace*{0cm}
\includegraphics[width=0.91\linewidth]{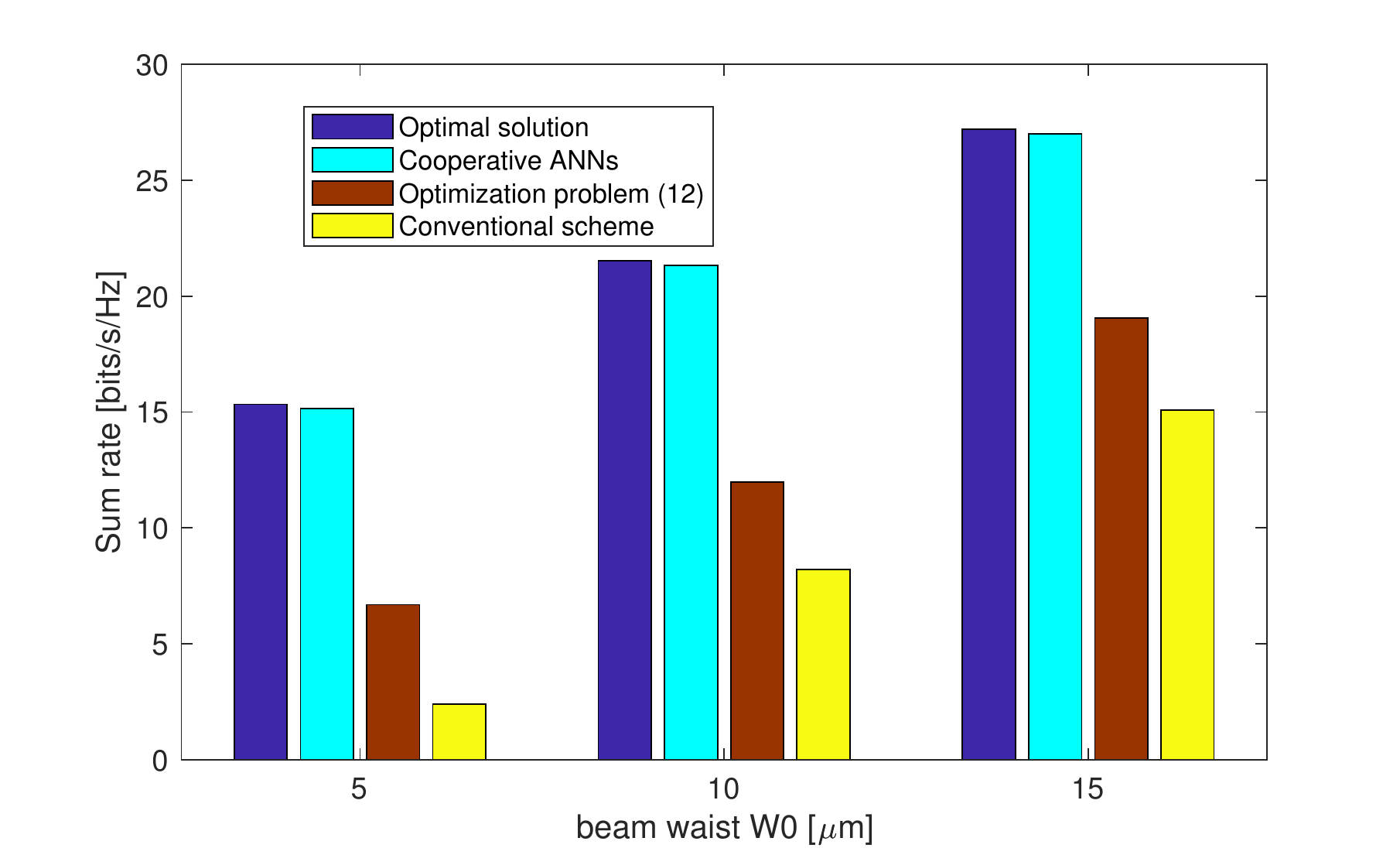}
\end{center}
\vspace{-2mm}
\caption{Sum rates of different rate-maximization techniques versus the beam waist of the laser. $\mathcal{T}=4$, $ K=10$. }\label{re2}
\vspace{-2mm}
\end{figure}

\section{PERFORMANCE EVALUATIONS}
\label{sec:Pcom}
We consider an indoor environment  with 5m$ \times $ 5m$  \times $ 3m dimensions where on the ceiling $ L=8 $   APs are deployed, and each AP with $ L_v $ transmitters. On the communication floor located at 2 m from the ceiling, $ K $ users are distributed randomly with different activities. Note that, each user $ k $ is equipped a reconfigurable detector that gives the opportunity to connect to most of the APs,  more details are in \cite{JLT1111}. All the other simulation parameters are listed in Table 1.
  
The accuracy of the trained ANN model, i.e., LSTM, in preforming prediction is depicted in Fig. \ref{re1} in terms of mean square error (MSE) versus a set of epochs. It is can be seen that training  and validation losses decrease with the number of epochs regardless of the dataset size considered since the optimal wights needed to preform specific mathematical calculations are set over time. However, increasing the size of dataset from 5000 to $ 10^4 $ results in decreasing the error of validation and training processes. It is worth noticing that MSE increases if more periods  of time, $ \mathcal{T}=5 $, is assumed for the same dataset size which is due to an increase in the prediction error. This issue can be avoided by training the ANN model over a larger dataset with data points $>10^4$ . The figure also shows that our ANN model is not overfitting and can predict accurate solutions in the online application where unexpected scenarios are more likely to occur. 

In Fig. \ref{re2}, the sum rate of the network is shown against different values of the beam waist $ W_0 $, which is known  as  a vital parameter in  laser-based OWC that influences the  power received at the user end. It is shown that the sum rate of the users increases with the beam waist due to the fact that more transmit power is directed towards the users, and less interference is received from the neighboring APs. Interestingly, our cooperative ANNs provides accurate solutions close to the optimal ones that involves high computational time. Note that, solving the 
the optimization problem in \eqref{pro-c2} results in low sum rates compared to our ANN-based solutions, which is expected due to the assumption of full connectivity, i.e., $    x^{[l,k]}_{t+1}=1 $, which in turn leads to wasting the resources. Moreover, the proposed models shows  superiority over the conventional scheme proposed in \cite{8636954} in which  each AP serves users located at a distance determining whether the signal received is useful or noise, and therefore, users are served regardless of their demands, the available resources and capacity limitations of the APs.
 
\begin{figure}[t]
\begin{center}\hspace*{0cm}
\includegraphics[width=0.91\linewidth]{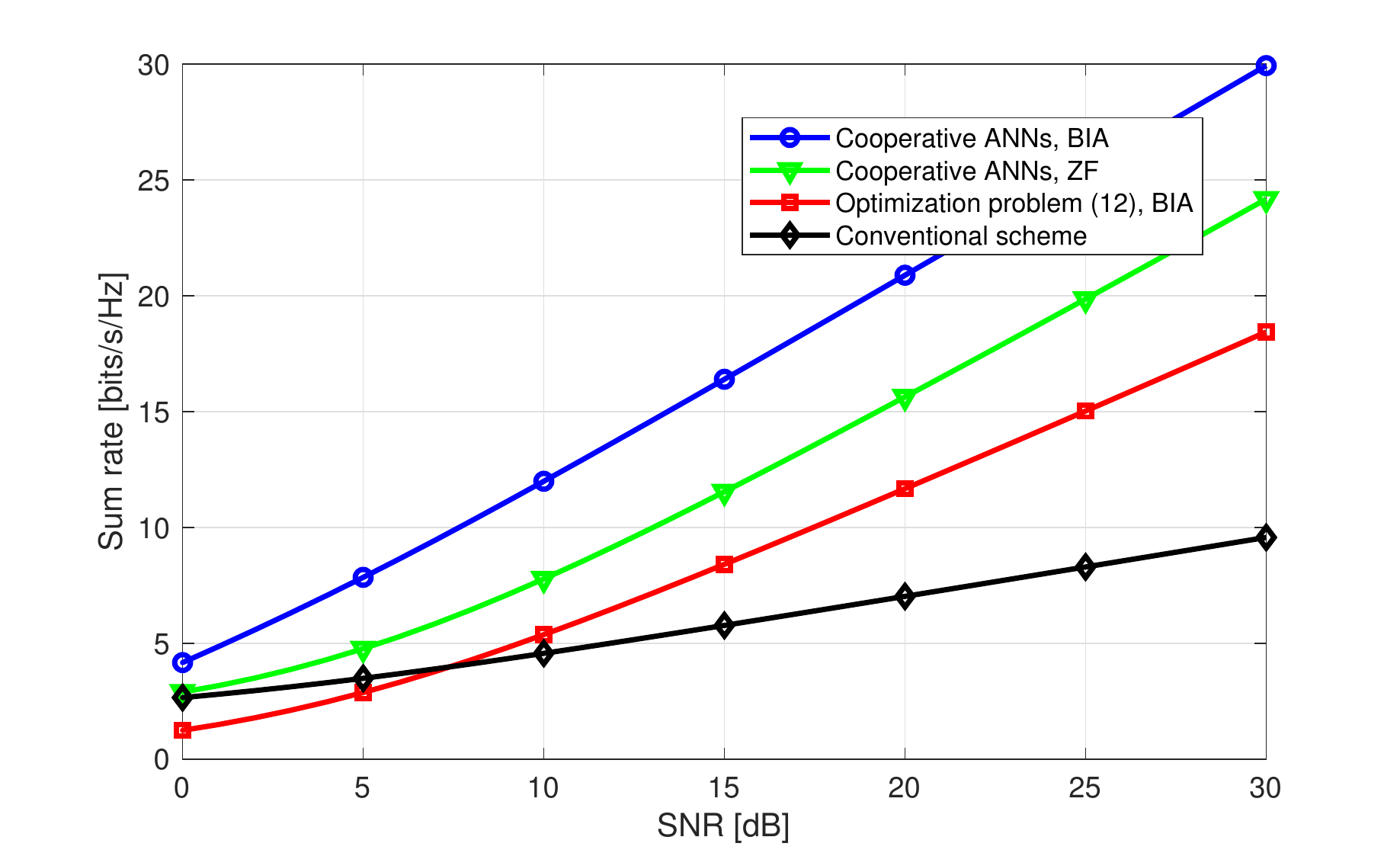}
\end{center}
\vspace{-2mm}
\caption{Sum rates of the network versus SNR. $\mathcal{T}=4$, $ K=10$.}\label{re3}
\vspace{-2mm}
\end{figure}

Fig. \ref{re3} shows the sum rate of the network  versus a range of SNR values  using the trained ANN models. It can be seen that determining the optimal user assignment and resource allocation using ANN results in higher sum rates compared to the scenarios  of full connectivity and distance-based user association. It is because of in our model, each user is assigned to an AP that  has enough resources to satisfy its demands and positively impact the sum rate of the network. Interestingly, as in \cite{8636954}, BIA  achieves a higher sum rate than ZF due to its ability to serve multiple users simultaneously with no CSI, while the performance of ZF  is dictated by the need for CSI.
 
\section{CONCLUSIONs}
\label{sec:CONCLUSION}
In this paper,  sum rate-maximization is addressed in a discrete time laser-based OWC. We first define the system model consisting of multiple APs to serve users distributed on the receiving plane. Then, the user rate is derived considering the application of BIA, which manages multi-user interference without the need for CSI at the transmitters. Moreover, an optimization problem is formulated to maximize the sum rate of the network during a certain period of time. Finally, CNN and LSTM models are designed and trained to provide instantaneous solutions during the validity of each period of time. The results show that solving the formulated model achieves higher sum rates compared to other benchmark models, and the trained ANN models have the ability to obtain accurate and valid solutions close to the optimal ones.

\bibliographystyle{IEEEtran}
\bibliography{IEEEabrv,mybib}

\end{document}